\documentclass[9pt,twocolumn,twoside]{osajnl}

\journal{preprint} 

\setboolean{shortarticle}{true}

\title{Ultra-broadband mid-infrared generation in dispersion-engineered thin-film lithium niobate}

\author[1,*]{Jatadhari Mishra}
\author[1,2]{Marc Jankowski}
\author[1]{Alexander Y. Hwang}
\author[1]{Hubert S.\ Stokowski}
\author[1,2]{Timothy P.\ McKenna}
\author[1]{Carsten Langrock}
\author[1,2]{Edwin Ng}
\author[1]{David Heydari}
\author[1]{Hideo Mabuchi}
\author[1]{Amir H.\ Safavi-Naeini}
\author[1,$\dagger$]{M. M.\ Fejer}

\affil[1]{E.\,L.\ Ginzton Laboratory, Stanford University, Stanford, California 94305, USA}
\affil[2]{NTT Research, Inc. Physics \& Informatics Laboratories, 940 Stewart Drive, Sunnyvale, California 94085, USA}

\affil[*]{email: jmishra@stanford.edu}
\affil[$\dagger$]{email: fejer@stanford.edu}

\ociscodes{(190.0190) Nonlinear Optics; (190.4360) Nonlinear optics, devices; (190.4390) Nonlinear optics, integrated optics.}

\begin{abstract}

Thin-film lithium niobate (TFLN) is an emerging platform for compact, low-power nonlinear-optical devices, and has been used extensively for near-infrared frequency conversion.
Recent work has extended these devices to mid-infrared wavelengths, where broadly tunable sources may be used for chemical sensing.
To this end, we demonstrate efficient and broadband difference frequency generation between a fixed 1-{\textmu}m pump and a tunable telecom source in uniformly-poled TFLN-on-sapphire by harnessing the dispersion-engineering available in tightly-confining waveguides. 
We show a simultaneous 1--2 order-of-magnitude improvement in conversion efficiency and $\sim$5-fold enhancement of operating bandwidth for mid-infrared generation when compared to conventional lithium niobate waveguides. 
We also examine the effects of mid-infrared loss from surface-adsorbed water on the performance of these devices.

\end{abstract}

\setboolean{displaycopyright}{true}

\begin{document}

\maketitle


The mid-infrared (MIR) window spanning 3--5~{\textmu}m contains the strongest absorption bands of many important gaseous molecules and functional groups, and thus is crucial for spectroscopic sensing and environmental monitoring~\cite{Vainio2016}.
Of the two primary approaches for MIR generation, semiconductor optoelectronics~\cite{Yao2012, Spott2016, Jung2017} and nonlinear optics~\cite{Schunemann2016, Petrov1998, Lind2020, Yu2018, Kowligy2018, Guo2020, Mishra2021}, the latter offers broader spectral coverage, wider tunability, and lower frequency noise~\cite{Foote2021, Borri2011} in the aforementioned spectral region. 
Nanophotonic implementations of nonlinear-optic MIR sources have received considerable attention in recent years, as they offer the additional benefits of compact integration, low power consumption, and deployment versatility. 
Most of these integrated photonic sources rely on the Kerr nonlinearity and on phase-velocity engineering~\cite{Yu2018, Kowligy2018, Guo2020}, but require relatively large input powers to achieve broadband operation. 
The recently developed periodically poled thin-film lithium niobate (TFLN) on sapphire platform, which avoids the problem of MIR absorption in the commonly used TFLN on silica substrates, leverages the stronger second-order ($\chi^{(2)}$) nonlinearity of LN for three-wave mixing with commonly available near-infrared (NIR) lasers to make possible a new generation of integrated MIR sources~\cite{Mishra2021}.
While this previous work showed a large normalized conversion efficiency for single-pass difference frequency generation (DFG), the DFG transfer functions in the 3-{\textmu}m band were relatively narrow ($\sim$20--30~nm), set by the dispersion of the interacting waves in these devices. 
However, since devices based on quasi-phasematched (QPM) $\chi^{(2)}$ interactions can realize efficient DFG for any waveguide geometry, the geometry can also be used as a design parameter to freely engineer the waveguide dispersion, and thus the bandwidth of the nonlinear interaction.
This technique of dispersion engineering has recently been been used for broadband second-harmonic generation, supercontinuum generation, and broadband parametric amplification in TFLN in the NIR~\cite{Jankowski2021, Ledezma2022, Jankowski2022}.

In this work, we employ dispersion engineering in periodically poled TFLN waveguides on sapphire substrates to design an ultra-broadband nanophotonic MIR source which can be driven by NIR tunable single-frequency sources, as well as NIR comb sources. 
We characterize this device via DFG between a fixed 1.064~{\textmu}m CW pump and a tunable CW telecom-band signal, and demonstrate MIR generation in the 2.8--3.8~{\textmu}m band with more than 700~nm (18.5 THz) FWHM bandwidth around 3.4~{\textmu}m in a single uniformly poled device and with a peak normalized efficiency of $\sim$100$\%$/W-cm$^2$.
Our device shows a 1--2 orders of magnitude higher normalized conversion efficiency when compared to other travelling-wave MIR DFG devices in large-core periodically poled LN (PPLN) ridge waveguides and diffused PPLN waveguides~\cite{Lehmann2019, Bchter2009, Mishra2021}, and about four orders of magnitude increase over bulk PPLN~\cite{Krzempek2014} (assuming 1-cm-long devices).
Additionally, the MIR bandwidth of our device is $\sim$5 times larger than previously reported for uniformly poled PPLN waveguides~\cite{Mayer2016}, and over 20 times larger than our previous demonstration in the TFLN platform~\cite{Mishra2021}. 
This is achieved without the need for the conventional chirped QPM approach that effectively reduces the interaction length for each wavelength and therefore increases the power requirements for a given nonlinear process~\cite{Lind2020}.
We also observe rapid tuning of the phasematching peak with temperature, at a fixed pump wavelength, with a rate $>$10~nm (300~GHz) per $^{\circ}$C or $\frac{{\Delta}\nu}{\nu}$ $\sim$3.2$\times$10$^{-3}/^{\circ}$C where $\nu$ is the frequency, which is one of the highest reported to date in this platform~\cite{Chen2020}.
Finally, we show that a dominant loss mechanism that can limit the performance of these devices is MIR absorption by surface-adsorbed water~\cite{Mishra2021}. 
When operated in a dry atmosphere and at elevated temperatures to desorb water from the surface, these devices exhibit the largest normalized efficiencies and bandwidths reported to date for MIR DFG.
We also discuss alternative solutions to the water adsorption problem.


\begin{figure}[!htb]
\centering
\includegraphics[width=\linewidth]{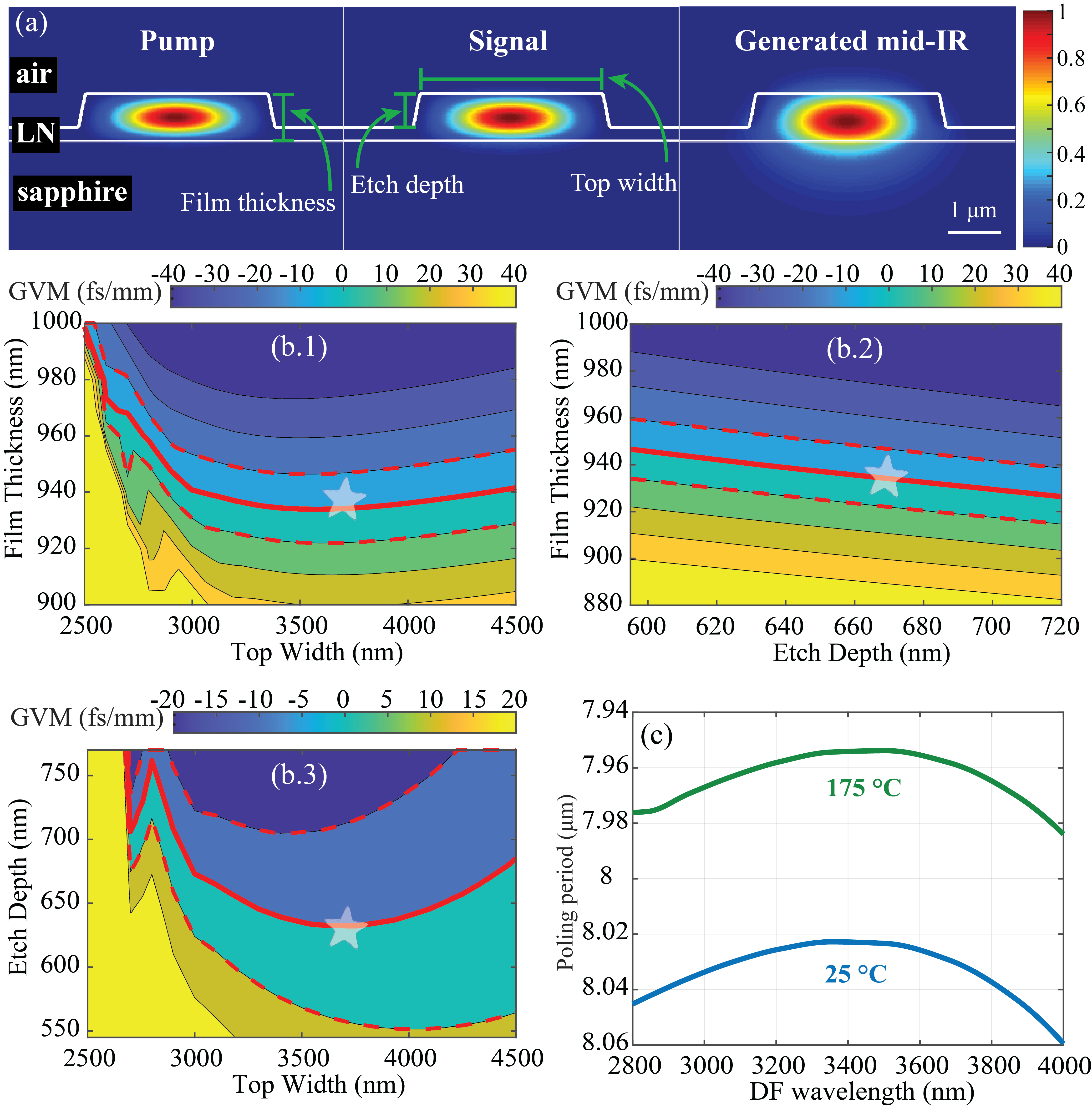}
\caption{
(a) Waveguide cross section and simulated fundamental TE mode intensity plots at wavelengths 1064~nm, 1529~nm, and 3500~nm, representing pump, signal, and idler, respectively. 
(b) Contour plots of simulated group-velocity mismatch (GVM) between signal and idler, at wavelengths 1560~nm and 3350~nm, respectively, for (b.1) varied top width and film thickness with etch depth fixed at 670 nm, (b.2) varied film thickness and etch depth with top width fixed at 3670 nm, (b.3) varied etch depth and top width with film thickness fixed at 940 nm. The zero-GVM contour is emphasized using solid red lines, while the |GVM| = 10 fs/mm contour is emphasized using dashed red lines. Star indicates the chosen geometry.
(c) Poling period vs generated difference frequency (DF) wavelengths (which is also the $\Delta k = 0$ contour) plotted for the geometry in (a) at two different temperatures.
}
\label{fig:design}
\end{figure}

We first discuss the design criteria for our dispersion-engineered MIR devices. 
As previously mentioned, we aim to generate coherent light (idler; $\omega_{i}$) over a large bandwidth in the MIR using the DFG process of a fixed 1064~nm CW pump ($\omega_{p}$) and a tunable CW telecom-band signal ($\omega_{s}$). 
The fundamental TE modes of all three waves (Fig.~\ref{fig:design}(a)) are phasematched via periodic poling of the TFLN-on-sapphire ridge waveguides, where sapphire is chosen as the substrate to allow low-loss transmission in the MIR~\cite{Mishra2021}. 
The bandwidth of a quasi-phasematched DFG process with a fixed pump is determined by the variation of the phase-mismatch $\Delta k(\Omega) = k(\omega_p) - k(\omega_s - \Omega) + k(\omega_i + \Omega) - 2\pi/\Lambda$ as the the input signal and thus the generated idler are detuned $\Omega$ relative to a nominally phase-matched frequency, $\omega_i$. 
Here, $k(\omega)$ is the propagation constant of the fundamental TE$_{00}$ mode, and $\Lambda$ is the poling period chosen to achieve $\Delta k(\Omega=0)=0$. 
To leading order in $\Omega$, the Taylor series expansion of the phase-mismatch is given by $\Delta k(\Omega) = (v^{-1}_{g,i} - v^{-1}_{g,s})\Omega$, where $v_{g,i/s} = (dk/d\omega_{i/s})^{-1}$ are the respective group velocities and ($v^{-1}_{g,i} - v^{-1}_{g,s}$) is referred to as the group-velocity mismatch (GVM) between the signal and idler waves~\cite{Jankowski2021}. 
Broadband operation occurs around wavelengths where the GVM $\approx$ 0.


The material dispersion causes the signal and idler waves to propagate at different group velocities and determines the interaction bandwidth. 
In nanophotonic waveguides however, the tight confinement of modal fields introduces a strong waveguide dispersion, which enables engineering of $\Delta k(\Omega)$ based on the choice of the waveguide geometry. 
During fabrication, we can independently control the TFLN film thickness, as well as the waveguide etch depth and top width~(Fig.~\ref{fig:design}(a)). 
Thus, in order to find the appropriate waveguide geometry for broadband DFG, we search through a three-dimensional parameter space, and visualize the GVM between the signal and the idler using a combination of multiple two-dimensional plots where two of the independent geometry parameters are swept, while the third is held constant.
Based on these parameter sweeps, we converged on an etch depth of $\sim$670~nm, top width of $\sim$3670~nm, and film thickness of $\sim$940~nm.
Figures~\ref{fig:design}(b.1, b.2 $\&$ b.3) show three of these two-dimensional GVM plots, where the etch depth, top width, and film thickness are held constant, respectively, at the aforementioned values, while the other two parameters are swept.
Apart from the zero-GVM contours, these plots also readily show the design tolerances based on the slopes of the said contours and contours of |GVM| = 10 fs/mm, which are together optimized to find the desired dispersion-engineered waveguide geometry.
The zeroing of the first-order term (GVM) in the Taylor series expansion of the phase-mismatch suggests a second-order (parabolic) dependence of phasematching on wavelength in these devices~\cite{Imeshev2000}.
This is shown in Fig.~\ref{fig:design}(c) for two different temperatures where the simulated phasematching contour ($\Delta k = 0$) is represented as the nominal required poling period.
We use the temperature-dependent LN dispersion model from Refs.~\cite{Umemura2014, Umemura2016} for all the above simulations.


Periodic poling of the thin-film and subsequent waveguide fabrication followed the procedure previously described in Ref.~\cite{Mishra2021} to achieve the designed geometry for all waveguides. 
A deeper etch depth than the previous work facilitated dispersion engineering and prevented lateral leakage of the short-wavelength TE$_{00}$ modes into slab modes for the chosen film thickness~\cite{Boes2019}.
The waveguides were $\sim$5-mm long, and designed with a 30-{\textmu}m-long non-adiabatic linear taper with a 500-nm top width at the input facet to facilitate fundamental-mode excitation of the pump and signal inputs.
While each waveguide was uniformly poled with a single poling period of around 8~{\textmu}m, the periods were swept across the device chip in steps of 20~nm in order to allow for wide temperature tuning of the phasematching peaks.


\begin{figure}[!htb]
\centering
\includegraphics[width=\linewidth]{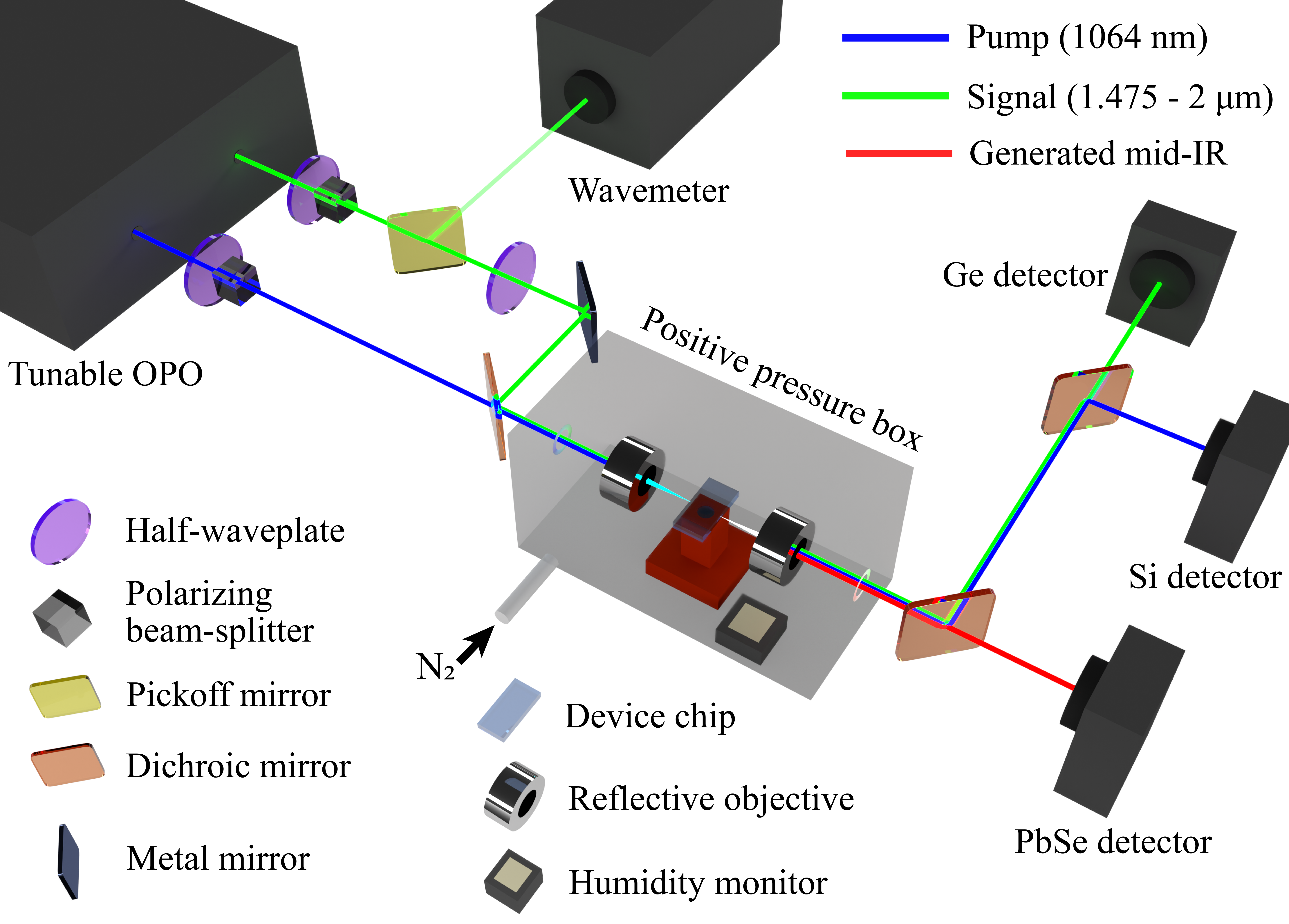}
\caption{Schematic of the experimental setup.}
\label{fig:experimental_layout}
\end{figure}

\begin{figure*}[!htb]
\centering
\includegraphics[width=\linewidth]{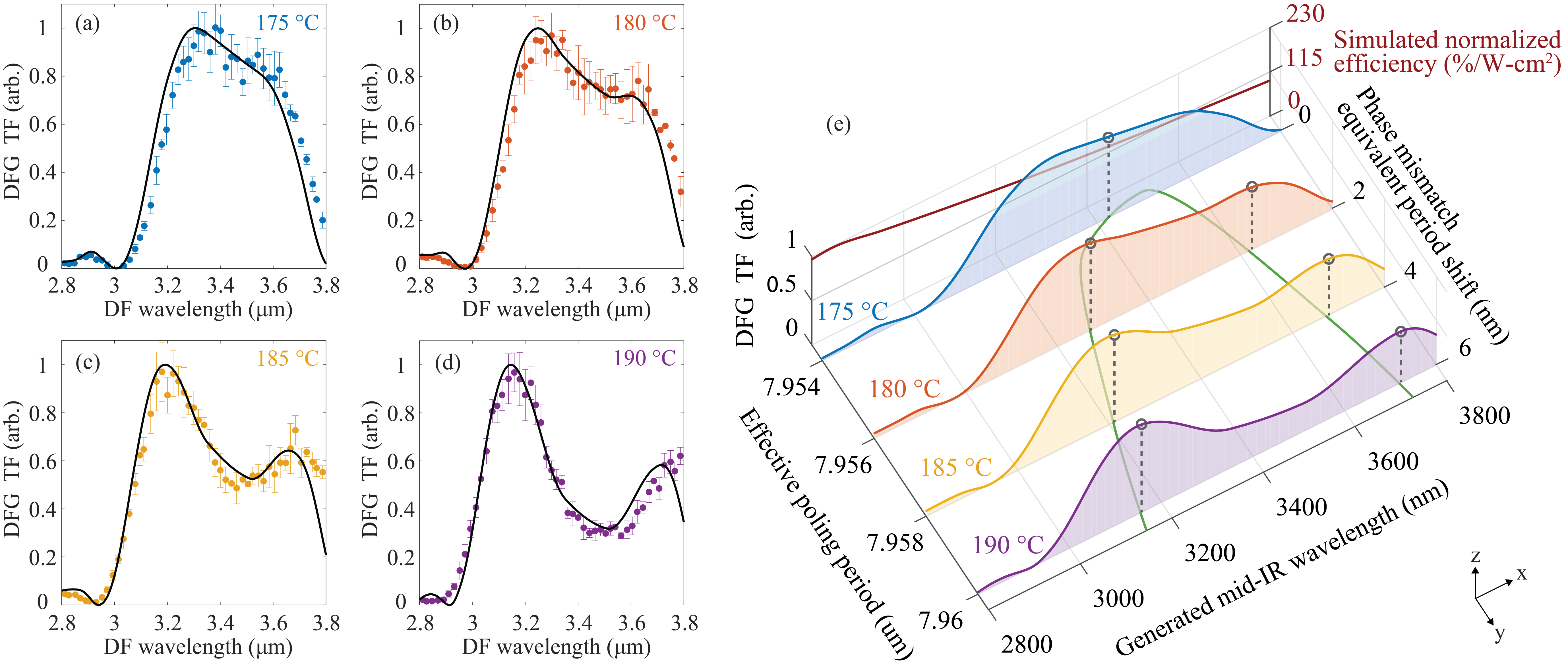}
\caption{(a--d) Measured (dots) vs simulated (solid lines) normalized DFG transfer functions (TF) at four different temperatures plotted against the difference-frequency (DF) wavelength. 
(e) Plotted against the DF / generated MIR wavelength (x-axis) are the individually normalized temperature-dependent simulated TFs from (a)--(d) in the xz-plane, the effective poling period (solid green) on the xy-plane, and the simulated wavelength-dependent normalized DFG efficiency (solid red) also in the xz-plane against a second z-axis.}
\label{fig:xfer_function_tuning}
\end{figure*}

The experimental setup is shown in Fig.~\ref{fig:experimental_layout}. A tunable CW optical parametric oscillator (OPO) (Toptica TOPO) was used as the source for both the fixed 1.064~{\textmu}m pump and the tunable (1.475--2~{\textmu}m) signal inputs to the waveguides. 
Appropriate half-wave plate and polarizing beam-splitter combinations served as variable optical attenuators for both input beams. 
A pickoff mirror in the path of the signal beam fed a wavemeter (Bristol 671), that continuously monitored the tuning of the OPO output signal wavelength. 
The signal beam passed through another half-wave plate, set to maintain TE polarization except during efficiency measurements as described later, and finally co-aligned with the pump beam via a dichroic mirror. 
The waveguide chip, mounted on a two-axis translation stage with vacuum chuck and temperature control, and both the input and output objectives (Thorlabs LMM-40X-P01) mounted on three-axis stages, were housed inside a positive pressure plexiglass box with nitrogen flow and cut-outs for the beams to enter and exit the box.
A humidity sensor was used to monitor the water content of the air inside the box. 
Reflective objectives were used to avoid chromatic aberrations during in and out coupling of the devices, enabling calibrated measurement of conversion efficiencies, though limiting input coupling efficiency to $\sim$ 1--2$\%$. Output collection efficiencies were $\sim25\%\pm1\%$ at 1~{\textmu}m, $\sim33\%\pm1\%$ in the 1.5-{\textmu}m band, and $\sim40\%\pm5\%$ in the 3-{\textmu}m band.
At the output of the setup, the transmitted pump and signal and generated MIR beams were separated using two dichroic mirrors and detected on a silicon power meter, a germanium power meter, and a lead selenide photoconductive detector (Thorlabs PDA20H), respectively. 
During alignment into a waveguide, the pump and signal modes at the output facet were imaged using silicon and InGaAs cameras respectively to maximize coupling to the respective TE$_{00}$ modes. 


We observed quasi-phasematched CW MIR DFG in the 2.8--3.8~{\textmu}m wavelength range in a single uniformly poled waveguide device by mixing the tunable signal beam (1.475--1.72-{\textmu}m) against the fixed pump beam (1.064-{\textmu}m).
Figs.~\ref{fig:xfer_function_tuning}(a-d) show the experimental (dots) and simulated (solid lines) DFG transfer functions plotted against the DFG wavelengths at four different temperatures (in the 175--190~$^{\circ}$C range) in the same waveguide (waveguide 1). 
We measured transfer functions with approximately 18.5 THz FWHM bandwidth ($\sim$700~nm) around a center wavelength of 3.4~{\textmu}m (Fig.~\ref{fig:xfer_function_tuning}(c)), and greater than 1~{\textmu}m or 26.5~THz bandwidth between the zeroes of the sinc (Fig.~\ref{fig:xfer_function_tuning}(d)). 
We observe excellent agreement between the measured and simulated transfer functions across many operating temperatures (175--190~~$^{\circ}$C) and note here that the only fitting parameter is a small constant offset of the phase-mismatch. 
To account for small changes, $\delta k$, of the phase-mismatch relative to the nominal design, we instead report an effective poling period shift $\Delta\Lambda=\Lambda^2\delta k/(2\pi)$.
While the lithographically patterned poling period for waveguide 1 is nominally 7.957~{\textmu}m at room temperature and expected to be $\sim$7.966~{\textmu}m at 175~$^{\circ}$C ($T_{0}$) owing to thermal expansion of the QPM grating, the measured transfer function obtained at the latter temperature is best fit with an effective poling period of 7.954~{\textmu}m, corresponding to $\Delta\Lambda_{0} = -12$~nm. 
This small period shift may be a result of slight deviations between the simulated and fabricated waveguide geometries, which are beyond the precision of the metrology tools used to characterize the waveguide.
We represent further contributions to the phase-mismatch ($\delta k(\Delta T)$) due to temperature tuning ($\Delta T = T - T_{0}$) as a temperature-dependent effective poling period shift ($\Delta\Lambda(\Delta T)$; Fig.~\ref{fig:xfer_function_tuning}(e), y-axis).
With the effective poling period fixed at its value chosen to fit at 175~$^{\circ}$C, simulations including both thermo-optic index changes and thermal expansion lead to results in agreement with the measured phase matching curves at the higher temperatures.
The simulated transfer functions $\eta(\omega_{i},T)$ at idler frequency $\omega_{i}$ and temperature T can thus be summarized as $\eta(\omega_{i},T)$ = $\eta_{0}(\omega_{i})$sinc$^{2}((\Delta k(\omega_{i}, T_{0}) + \delta k(\Delta T))L/2)$, where $L$ is the device length, and $\eta_{0}(\omega_{i}) \propto \omega_{i}^{2}$ is the wavelength dependent normalized efficiency~\cite{Jankowski2021} in units of $\%$/W-cm$^2$ (Fig.~\ref{fig:xfer_function_tuning}(e), solid red curve).
We note that in Fig.~\ref{fig:xfer_function_tuning}(e) the phasematching peaks at 180, 185, and 190~$^{\circ}$C line up well with the poling period vs DFG wavelength curve simulated in Fig.~\ref{fig:design}(c).
However, the position of the peak of the transfer function at 175~$^{\circ}$C deviates from the wavelength at which we have zero phase-mismatch due to the wavelength dependence of the normalized efficiency and the relative flatness of the sinc$^{2}(\Delta k)$ term.

\begin{figure}[!htb]
\centering
\includegraphics[width=\linewidth]{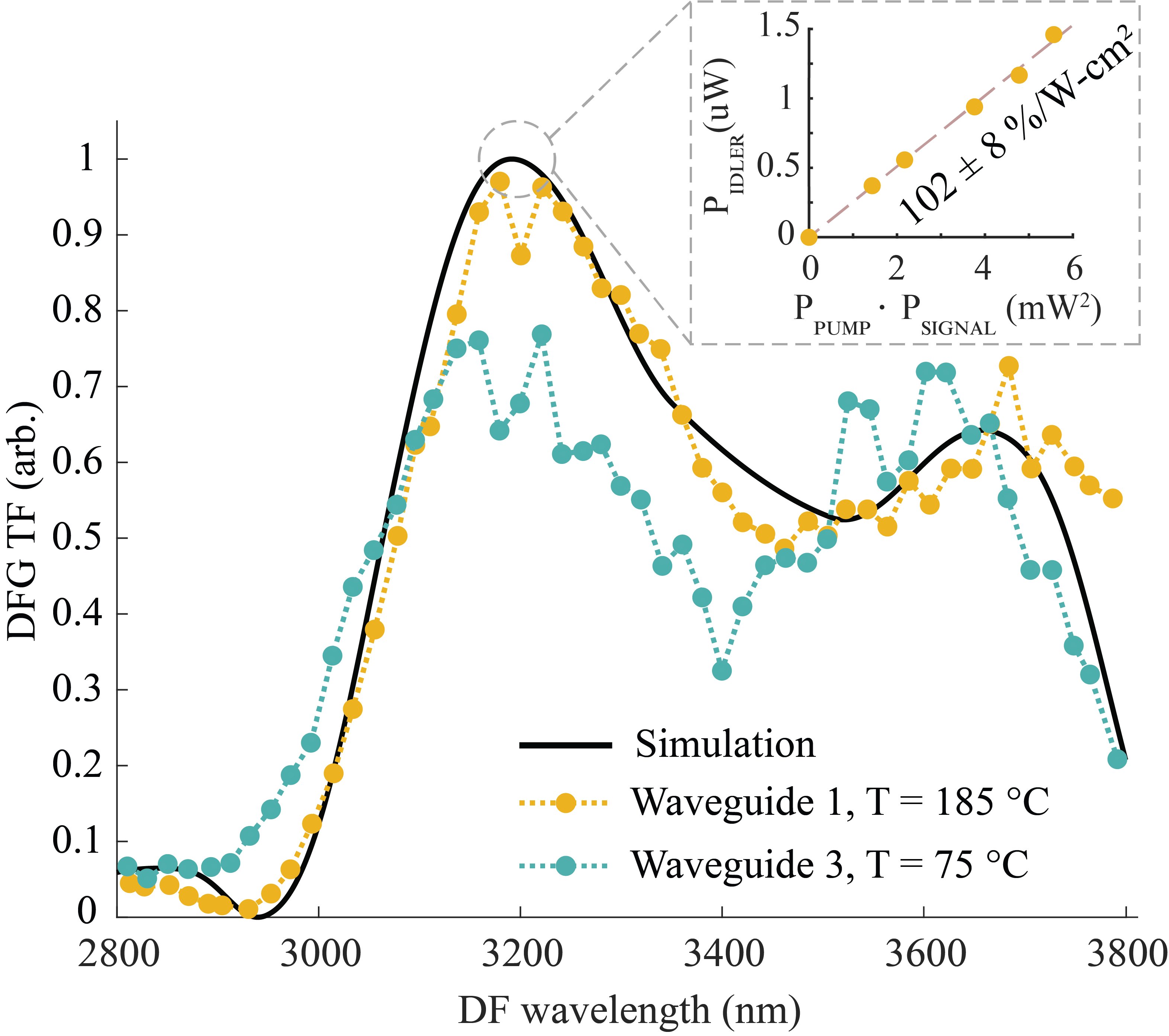}
\caption{Measured transfer functions (dashed-dots) for waveguide 1 at 185~$^{\circ}$C and waveguide 3 at 75~$^{\circ}$C. 
Waveguide 3 has a nominal poling period of +40~nm relative to waveguide 1. 
Solid curve is the simulated transfer function.
Both the simulation and waveguide 1 transfer functions are normalized to 1, while the waveguide 3 transfer function is re-scaled to approximately match the height around 3.6--3.7~{\textmu}m with that of waveguide 1. 
Inset shows change of generated MIR (idler) power with change of signal power in the TE polarization when pump power is held constant.}
\label{fig:normalized_efficiency}
\end{figure}

We also observed rapid tuning of the phasematching peak with temperature, with a value greater than 300~GHz (10~nm at the idler wavelength) per $^{\circ}$C ($\frac{{\Delta}\nu}{\nu}$ $\sim$3.2$\times$10$^{-3}/^{\circ}$C).
This temperature tuning of the phasematching allowed us to observe the same nonlinear interaction in the neighboring waveguide (20~nm nominal poling period difference) with every $\sim$55~$^{\circ}$C change in temperature of the device. 
However, while the higher temperature range (>150~$^{\circ}$C) measurements shown in Fig.~\ref{fig:xfer_function_tuning} are in good agreement with simulations, the shapes of the transfer functions measured at lower temperatures deviated significantly from simulations (Fig.~\ref{fig:normalized_efficiency}, waveguide 1 vs waveguide 3). 
We believe that the reason for this deviation at lower temperatures is wavelength-dependent propagation loss in the MIR due to absorption by water adsorbed on the air-clad surfaces of the TFLN waveguides, which reduces the measured DFG efficiency relative to theory.
When operated in a dry atmosphere (<10$\%$ humidity) and at high temperature, the adsorbed water is driven off, restoring the expected transfer function. 
The adsorbed-water-dependent propagation loss is estimated to be the strongest around 2.8~{\textmu}m and weaker at longer wavelengths~\cite{Mishra2021}.
To examine the effect of this wavelength-dependent loss on the observed DFG, we rescale the spectrum obtained at 75 $^{\circ}$C (where water is adsorbed on the surface) so that its long-wavelength peak (where the loss is weak) matches that of the spectrum obtained at 185 $^{\circ}$C (where the water is largely desorbed). 
The difference of $\sim$25$\%$ in the short-wavelength peaks (where the loss is strong) is comparable  to that expected from the loss estimates in \cite{Mishra2021}. 

The normalized efficiency was measured at the phasematching peak at 185~$^{\circ}$C. 
For this measurement, the pump and signal wavelengths, as well as their powers, were held constant. 
Only the content of TE-polarization in the signal beam was changed by rotating the second half-wave plate in its path (Fig.~\ref{fig:experimental_layout}). 
The generated DFG power was observed to be linearly dependent (Fig.~\ref{fig:normalized_efficiency} inset) on the signal TE power, where the latter was detected by using a linear polarizer before the Ge power meter. 
The detected powers, in combination with the respective collection efficiencies for the output objective, were used to derive the amount of power at each wavelength inside the waveguide. 
A normalized DFG efficiency of 102$\pm$8$\%$/W-cm$^2$ was calculated in the low-conversion limit by taking the ratio of the on-chip MIR power (at the output) over the product of the on-chip pump and signal (TE) powers. 
While we have provided the standard deviation in our measurements, we note that the measured efficiency can only be considered accurate within a few tens of percent, owing to uncertainties in the coupling and detector calibrations, particularly in the MIR. 
The simulated DFG normalized efficiency at this wavelength is 167$\%$/W-cm$^2$. 
Difference between simulated and experimental efficiencies can be attributed to imperfections in poling and waveguide fabrication.

In conclusion, we have demonstrated a nanophotonic ultra-broadband NIR to MIR wavelength converter in the TFLN-on-sapphire platform that uses dispersion engineering to achieve a $\sim$5 times larger DFG bandwidth than previous state-of-the-art PPLN-based single-pass MIR generators in the 3--4~{\textmu}m band.
The TFLN device also benefits from the sub-wavelength mode confinement to achieve a 1--2 orders-of-magnitude higher internal normalized efficiency compared to the conventional LN waveguide devices. 
The generated MIR light is tuned within the large phasematching bandwidth by changing the frequency of a telecom-band signal, while the 1.064~{\textmu}m pump is fixed. 
By changing the device temperature, we could tune the phasematching peak itself over the entire 2.6--4{\textmu}m MIR window and possibly wider, at a rate that is one of the highest reported in this platform. 
We also established that the previously observed MIR loss in these waveguides was due to adsorbed water at the air-clad surfaces of the TFLN waveguides.
The demonstrated low power requirements and broadband tuning range open up a number of possibilities, including integrated systems where NIR semiconductor lasers can be used to generated MIR tunable sources, and intrapulse DFG systems where an NIR pulse is used to form a carrier-envelope-phase stable MIR comb.
Suitable top cladding to eliminate adsorbed-water absorption may also enable low-loss operation at room temperature.

\begin{backmatter}
\bmsection{Funding} The authors wish to thank NTT Research for their financial and technical support through award 146395. This work was also supported by the NSF (CCF-1918549, PHY-2011363), the DARPA Young Faculty Award (DARPA-RA-18-02-YFA-ES-578), and the Department of Energy (DE-AC02-76SF00515). Devices were fabricated at the Stanford Nanofabrication Facility (NSF ECCS-1542152), the Stanford Nano Shared Facilities (NSF ECCS-2026822), and the Cell Sciences Imaging Facility (NCRR S10RR02557401). H.S.S. acknowledges support from the Urbanek Family Fellowship. A.Y.H. is supported by the NSF Graduate Research Fellowship, Grant No. 214675.

\bmsection{Disclosures} The authors declare no conflicts of interest.
\bmsection{Data Availability Statement} Raw data underlying the results presented in this paper may be obtained from the authors upon reasonable request.
\end{backmatter}

\bibliography{references}

\bibliographyfullrefs{references}

\end{document}